\documentstyle[aps,prl,twocolumn]{revtex}

\begin{document}

\draft

\title{Ground State of a Homogeneous Bose Gas: A Diffusion Monte Carlo Calculation}

\author{S. Giorgini$^{1}$, J. Boronat$^{2}$ and J. Casulleras$^{2}$}

\address{$^{1}$Dipartimento di Fisica, Universit\`a di Trento, \protect\\
and Istituto Nazionale di Fisica della Materia, I-38050 Povo, Italy}
\address{$^{2}$Departament de F\'{\i}sica i Enginyeria Nuclear, Campus Nord B4-B5, 
\protect\\ Universitat Polit\`ecnica de Catalunya, E-08034 Barcelona, Spain}


\maketitle

\begin{abstract}

{\it We use a diffusion Monte Carlo method to calculate the lowest energy state of a uniform gas of 
bosons interacting through different model potentials, both strictly repulsive and with an attractive
well. We explicitly verify that at low density the energy per particle follow a universal behavior 
fixed by the gas parameter $na^3$. In the regime of densities typical for experiments in trapped
Bose-condensed gases the corrections to the mean-field energy greatly exceed the differences 
due to the details of the potential.}

\end{abstract}


\narrowtext

The achievement of Bose-Einstein condensation (BEC) in magnetically trapped atomic vapours
\cite{EXP} has revived interest in the theoretical study of Bose gases.
Mean-field methods provide us with relatively simple predictions both for the equilibrium properties 
of these systems (energy per particle, density profiles, condensate fraction) and for the dynamic 
behavior (frequency of collective excitations, interference effects), which have been 
found in close agreement with experiments (for a review see Ref. \cite{REV}).  
In fact, the atomic clouds realized in experiments are very dilute, the average distance 
between particles being significantly larger than the range of interatomic forces, 
and mean-field approaches are well suited. However, the investigation of effects beyond 
mean-field theory is an important task, which would make these systems even more interesting from 
the point of view of many-body physics. Theoretical studies of these effects have already 
been proposed, either by analytic inclusion of fluctuations around mean-field \cite{BMFAN,PS} 
or through numerical calculations based on quantum Monte Carlo methods \cite{BMFMC} and, 
more recently, also on correlated basis function approaches \cite{FP}.
All these investigations are based on the idea that, for the values of density relevant in 
experiments, the details of the interatomic potential can be neglected and one can safely use 
the hard-sphere model in numerical simulations, and the expansion in powers of the gas parameter 
$na^3$, fixed by the number density $n$ and the {\it s}-wave scattering length $a$, in analytic corrections 
beyond mean-field. 
The main motivation of the present work is to verify the validity of this approach.
By using a diffusion Monte Carlo (DMC) method we calculate the ground-state energy of a system of 
bosons interacting through different two-body model potentials. We explicitly show that for the values 
of the gas parameter reached in magnetic traps ($na^3 \simeq$ 1E-5 - 1E-4) the behavior is universal 
and fixed by $na^3$ and that the corrections to the mean-field energy are much larger than the 
differences due to the details of the interatomic potential. 

The ground state of a homogeneous dilute Bose gas was intensively studied in the 50's 
and early 60's. One of the main results of this investigation is that the  
ground-state energy can be expanded in powers of $\sqrt{na^3}$.
In units of $\hbar^2/2ma^2$ the energy per particle takes the form
\begin{eqnarray}
\frac{E}{N} = 4\pi na^3 \Bigg[ 1 &+& \frac{128}{15\sqrt{\pi}}\sqrt{na^3} 
\nonumber \\
&+& \frac{8(4\pi-3\sqrt{3})}{3}na^3 \ln(na^3) + ... \Bigg] \;.
\label{enexp}
\end{eqnarray}  
The first term corresponds to the mean-field prediction and was already calculated by 
Bogoliubov \cite{BOG}. The corrections to mean-field have been obtained using perturbation 
theory: the coefficient of the $(na^3)^{3/2}$ term was first calculated by Lee, Huang 
and Yang \cite{LHY}, while the coefficient of the last term was first obtained by Wu \cite{WU}.
Both of them were originally derived for hard spheres, but it was shown that the 
same expansion is valid for any repulsive potential with scattering length $a$ 
\cite{BSBL,HP}. Furthermore, Hugenholtz and Pines \cite{HP} have shown that the higher-order 
terms in the expansion (\ref{enexp}) depend on the ``shape'' of the interatomic potential.
It is worth pointing out that, for the values of $na^3$ relevant in experiments, the corrections to 
the mean-field energy are very small ($\simeq$ 3\%).
In a recent paper Lieb and Yngvason \cite{LY} have also provided a rigorous lower bound for the 
ground-state energy of a Bose gas holding for non-negative, finite range, spherical, two-body 
potentials. The lower bound coincides with the Bogoliubov mean-field term: $E/N\ge 4\pi(na^3) 
\hbar/2ma^2$. However, what is not well established up to now is the range of validity of the 
universal law (\ref{enexp}). 

In realistic systems the above restrictions for the interatomic potential do not hold, because 
an attractive tail is generally present. In this case the situation for the ground state is 
different. If the potential does not sustain any many-body bound state and the scattering length
is positive, the ground 
state of the system still behaves like a gas and the expansion (\ref{enexp}) should hold. 
Conversely, in the case of potentials which have two, three or more-body bound states, the 
ground-state of the system is no longer a homogeneous gas, but a state with clusters of atoms 
formed. However, if the scattering length is positive, a uniform gas can still exist as a metastable 
state which will be long-lived at very low densities. Our results for a simple model of interatomic 
potential with an attractive part show that the energy of the gas-like metastable state still follows 
the universal law given by (\ref{enexp}).

We consider a system of $N$ spinless bosons with mass $m$ described by the many-body Hamiltonian
$\hat{H}=- (\hbar^2/2m)\sum_{i=1}^N\nabla_i^2+\sum_{i<j}V(|{\bf r}_i-{\bf r}_j|)$,
where $V(r)$ is the two-body, spherical, interatomic potential.
The DMC algorithm solves exactly, apart from statistical uncertainty, the $N$-body Schr\"odinger 
equation for the ground-state energy of the system (for further details on the method see for 
example Ref. \cite{US}). We have used different choices for the potential $V(r)$:

\noindent
1) Hard-sphere (HS) potential defined by
\begin{equation}
V(r)=\left\{  \begin{array}{cc} +\infty & (r<a)  \\
                                   0    & (r>a)  \end{array} \right.
\label{HS}
\end{equation}
where the diameter $a$ of the hard sphere corresponds to the scattering length.

\noindent
2) Soft-sphere (SS) potential defined by    
\begin{equation}
V(r)=\left\{  \begin{array}{cc}    V_0  & (r<R)  \\
                                   0    & (r>R)  \end{array} \right.
\label{SS}
\end{equation}
with a scattering length: 
$a=R[1-\tanh(K_0R)/K_0R]$ with $K_0^2=V_0m/\hbar^2$ and $V_0>0$. 
For finite $V_0$ one has always $R>a$, while for $V_0\to+\infty$ the SS potential coincides with 
the HS one with $R=a$. For the SS potential we have considered two choices: $R=5a$ and $R=10a$.
It is worth noticing that the hard sphere and the very soft sphere with $R=10a$ represent two 
extreme cases for a repulsive potential. In the HS case, the energy is entirely kinetic, while for the 
very ``soft'' potential $a\simeq(m/\hbar^2)\int_0^{\infty}V(r)r^2 dr$, according to Born approximation,
and the energy is almost all potential. In the latter case the ground-state wave function is 
close to the noninteracting one (see Fig. 4).
 
\noindent
3) Hard-core square-well (HCSW) potential defined by
\begin{equation}
V(r)=\left\{  \begin{array}{cc}    +\infty  & (r<R_c)   \\
                                   -V_0     & (R_c<r<R) \\
                                    0       & (r>R)     \end{array} \right.
\label{HCSW}
\end{equation}
for which the {\it s}-wave scattering length is given by: 
$a=R_c+(R-R_c)\{1-\tan[K_0(R-R_c)]/K_0(R-R_c)\}$
with $K_0^2=V_0m/\hbar^2$ and $V_0>0$. For fixed $R_c$ and $R$ the scattering length coincides 
with the hard-core radius $R_c$ for $V_0=0$ and by increasing $V_0$ it exhibits resonances
each time a two-body bound state appears in the well. In many real potentials (such as in $^{87}$Rb 
and $^{23}$Na) the scattering length is much larger than the size of the atom. To reproduce 
this situation we have chosen $R=5R_c$ and $a=10R$ with only one two-body bound state in the well.

In the case of purely repulsive potentials we apply directly the DMC algorithm to obtain the ground-state 
energy. As importance sampling we use a Jastrow trial function: $\psi_T({\bf R}) \equiv 
\psi_T({\bf r}_1,..,{\bf r}_N) = \prod_{i<j}f(r_{ij})$. The Jastrow factor $f(r)$ is chosen as the 
exact wave function of a pair of particles interacting through $V(r)$ with total energy $\epsilon$ 
for $r\le\bar{R}$, and for $r>\bar{R}$ a function $f(r)=1-Ae^{-r/\alpha}$ which goes rapidly to one. 
The parameter $\alpha$ is left as a variational parameter, while the coefficient $A$ and the matching
point $\bar{R}$ are chosen so that $f(r)$ and its first derivative be continous at $r=\bar{R}$ and the 
local energy $(-\hbar^2\nabla^2/2m+V)f(r)/f(r)$ be also continous at $r=\bar{R}$. The energy 
$\epsilon$ is the second variational parameter. Before starting the DMC calculation, we perform  
a variational Monte Carlo (VMC) analysis to optimize the parameters of the trial wave function.
At low densities, we find that the DMC calculation improves very little on the VMC result. For 
example, for the HS potential at $na^3=$1E-5 and in units of $\hbar^2/2ma^2$: 
$E_{VMC}/N=$1.278(1)E-4 and $E_{DMC}/N=$1.274(1)E-4. This means that at low densities 
$\psi_T({\bf R})$ has a large overlap with the ``true'' ground-state wave function.

The case of the HCSW potential needs a careful treatment. Since the potential has a two-body 
bound state, the gas-like state is not the ground state. To obtain the energy of the metastable 
gas-like state it is necessary to project out the bound-state component of the wave function. 
This can be achieved by using the same trial wave function as for the SS potential 
(any trial function which is positive in the region where the potential 
is attractive would be equally appropriate) and then projecting the results for the energy by 
means of an auxiliary function $\psi_P({\bf R})$ ortogonal to any many-body state with bound 
pairs. In the Monte Carlo formulation this is realized by the weighted integral 
$E=\int d{\bf R} f({\bf R},\tau\to\infty) [\hat{H}\psi_P({\bf R})]/ \psi_T({\bf R})/
\int d{\bf R} f({\bf R},\tau\to\infty) \psi_P({\bf R})/\psi_T({\bf R})$, where 
$f({\bf R},\tau\to\infty)$ is the density of walkers generated by the DMC calculation. 
The projecting
function $\psi_P({\bf R})$ is chosen as: $\psi_P({\bf R})=\prod_{i<j}f_P(r_{ij})$, where $f_P(r)$
coincides with the trial two-body function $f(r)$ for $r>\bar{R}$, while for $r<\bar{R}$ is given
by the exact solution for two particles interacting through $V(r)$ with energy $\epsilon>0$ 
(we require the same continuity conditions at the matching point $\bar{R}$ as for $f(r)$). 
Since the matching point $\bar{R}$ is much larger than the range $R$ of the potential, $f_P(r)$ is 
orthogonal to the bound-state wave function, and $\psi_P({\bf R})$ is orthogonal to any many-body 
state with one or more bound pairs formed. In this way, we eliminate from the calculation all states 
with bound pairs. At low densities we expect that bound pairs give the main contribution to the ground
state compared to three or more-body bound states. By eliminating these bound pairs we are thus left
with the gas-like metastable state with lowest energy.

We are now in a position to discuss our results. In Table I we present the DMC results for the 
equation of state of the HS potential. In all the calculations we have used a simulation box 
containing $N=500$ particles in order to reduce finite size effects well below statistical 
uncertainty. Previous Monte Carlo calculations of the ground-state 
energy of the hard-sphere boson system have been performed at densities typical of liquid $^4$He 
($na^3\simeq 0.1$) \cite{HLS,KLV}. The results for the two highest densities in Table I are 
compatible with the Green's Function Monte Carlo results of Ref. \cite{KLV}. In Fig. 1, our 
results for the HS equation of state are shown together with the various terms of the analytic 
expansion (\ref{enexp}). One can see that the Lee-Huang-Yang (LHY) correction [second term in
(\ref{enexp})] represents a significant improvement on the mean-field prediction and the 
inclusion of this term allows for a good approximation of the equation of state up to very high
densities. On the contrary, the logarithmic correction [third term in (\ref{enexp})] goes wrong 
already at intermediate densities ($na^3 \simeq$ 1E-3).

In Table II, we report the results of the comparison between the various potentials considered 
in this work (the corresponding results for the HS potential can be read from Table I).
At low values of the gas parameter $na^3$ all potentials give the same results and only at 
the largest density reported ($na^3=$1E-3) the results for the SS potential start to deviate from 
the HS values. The universal behavior
is better shown in Fig. 2, where we plot the difference between the calculated energy per particle 
and the mean-field term and compare it with the LHY correction. It is worth mentioning that this 
difference is always positive in agreement with the lower bound provided by the mean-field term, 
as discussed in Ref. \cite{LY}.
The very low density regime and the relevance of the logarithmic correction is analysed in the 
inset of Fig. 2.
The HS and SS results show evidence of the presence of this logarithmic correction, however the 
effect is tiny and goes rapidly wrong for larger densities. Due to larger error bars no conclusion
can be drawn for the HCSW results concerning the logarithmic term.  

Another quantity of interest is the condensate fraction $N_0/N$ which we calculate from the 
long-range behavior of the one-body density matrix: $N_0/N=\lim_{r\to\infty}\rho(r)$ (see Ref. 
\cite{US} for further details). In Fig. 3, we show the results for $N_0/N$ as a function
of $na^3$ for the HS and the two SS potentials using the extrapolated estimator. 
The results are compared with the analytic expansion
\begin{equation}
\frac{N_0}{N}=1-\frac{8}{3\sqrt{\pi}} (na^3)^{1/2} \;,
\label{cfrac}
\end{equation}
calculated by Bogoliubov \cite{BOG}. At low densities this law is universal and agrees very well 
with the results of the three model potentials. First deviations from universality start to appear 
for $na^3\simeq$ 1E-3, result which coincides with the emergence of deviations in the energy values. 

In Fig. 4 we show the two-body distribution function $g(r)$ for hard spheres at various densities
obtained from the DMC calculation using the method for pure estimators of Ref. \cite{US1}. 
One can clearly see the building up of short range correlations at high 
densities. The distribution function $g(r)$ can be estimated at low densities using the Bogoliubov
result for the static form factor: $S(k)=\hbar^2k^2/2m\epsilon(k)$, where 
$\epsilon(k)=(\hbar^4k^4/4m^2+gn\hbar^2k^2/m)^{1/2}$ is the usual Bogoliubov dispersion relation 
with coupling constant $g=4\pi\hbar^2a/m$.The Fourier transform of $S(k)$ gives directly the 
distribution function $g(r)$, which for $r\gg a$ exhibits the asymptotic behavior 
$g(r)\simeq 1-[4\pi^{5/2}(na^3)^{3/2}(r/a)^4]^{-1}$.
In Fig. 4 we also compare, at the density $na^3=$1E-4, the distribution function obtained from the 
Bogoliubov $S(k)$ with the $g(r)$ of hard spheres and soft spheres with $R=10a$.  
The long range behavior is universal and agrees with the Bogoliubov result, while at short distances 
the HS and SS potentials give completely different results since in the second case particles can 
penetrate the repulsive potential. At higher densities the Bogoliubov approximation for $g(r)$ 
becomes poorer and misses completely the shell structure of the distribution function.

In conclusion, we have shown that for the values of density realized in magnetic traps the 
corrections to mean-field are fixed only by the gas parameter $na^3$ and do not depend on the 
details of the interatomic potential. These effects, although small, have a simple 
theoretical description in terms of the parameter $na^3$ and, possibly, can be singled out in 
future precision measurements \cite{PS}.

The authors would like to thank S. Stringari and L.P. Pitaevskii for many useful discussions.
This research has been partially supported by DGES (Spain) Grant n$^0$ PB96-0170-C03-02.

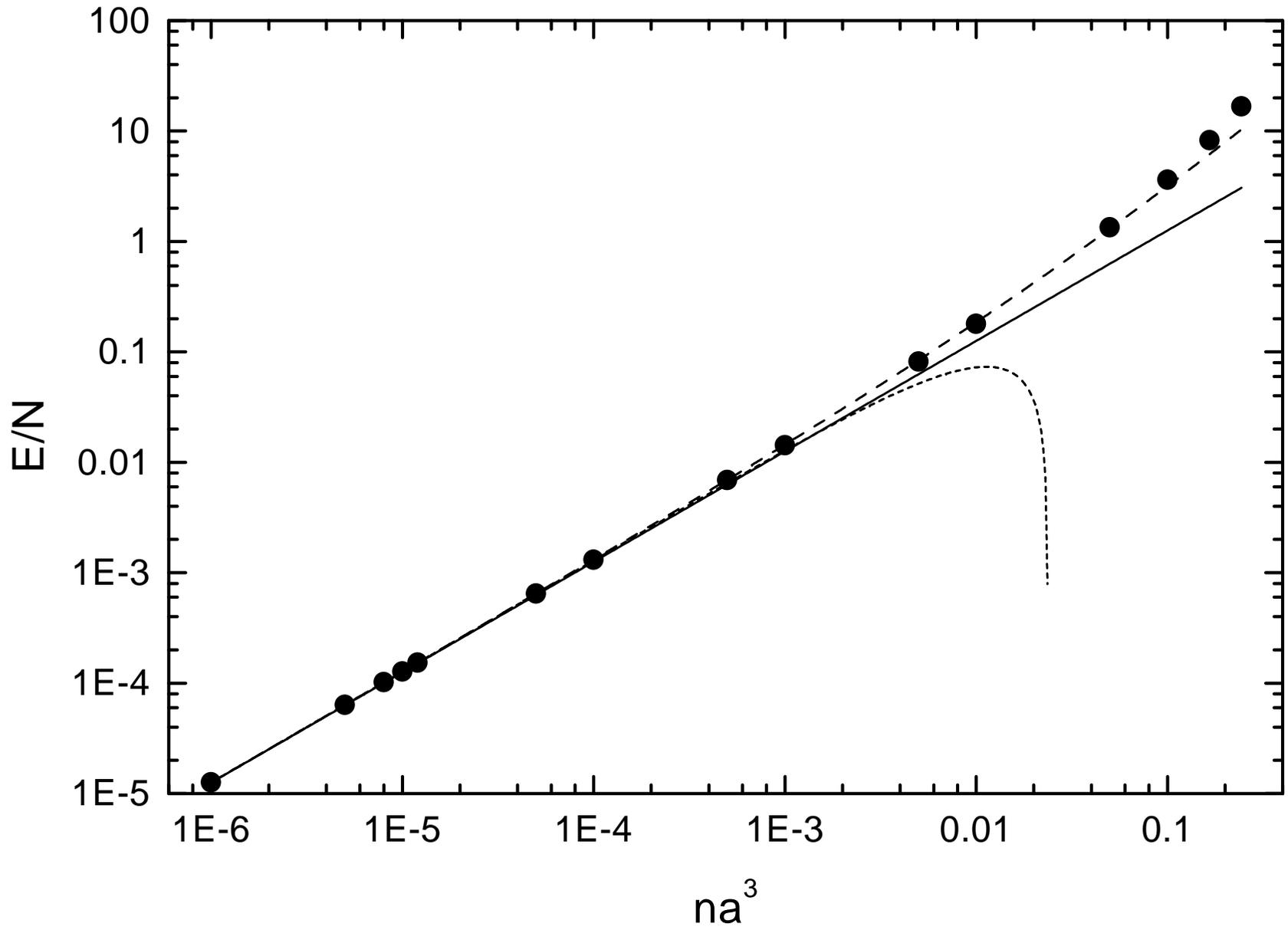
\begin{figure}
\caption{Equation of state for the HS potential. The solid circles are the DMC energies (error bars
are smaller than the size of the symbols); the solid line corresponds to the mean-field prediction 
[first term in eq. (\ref{enexp})]; the long-dashed line includes the LHY correction [first two terms
in eq. (\ref{enexp})]; the short-dashed line includes also the logarithmic correction and corresponds
to the full expansion (\ref{enexp}).}
\end{figure}

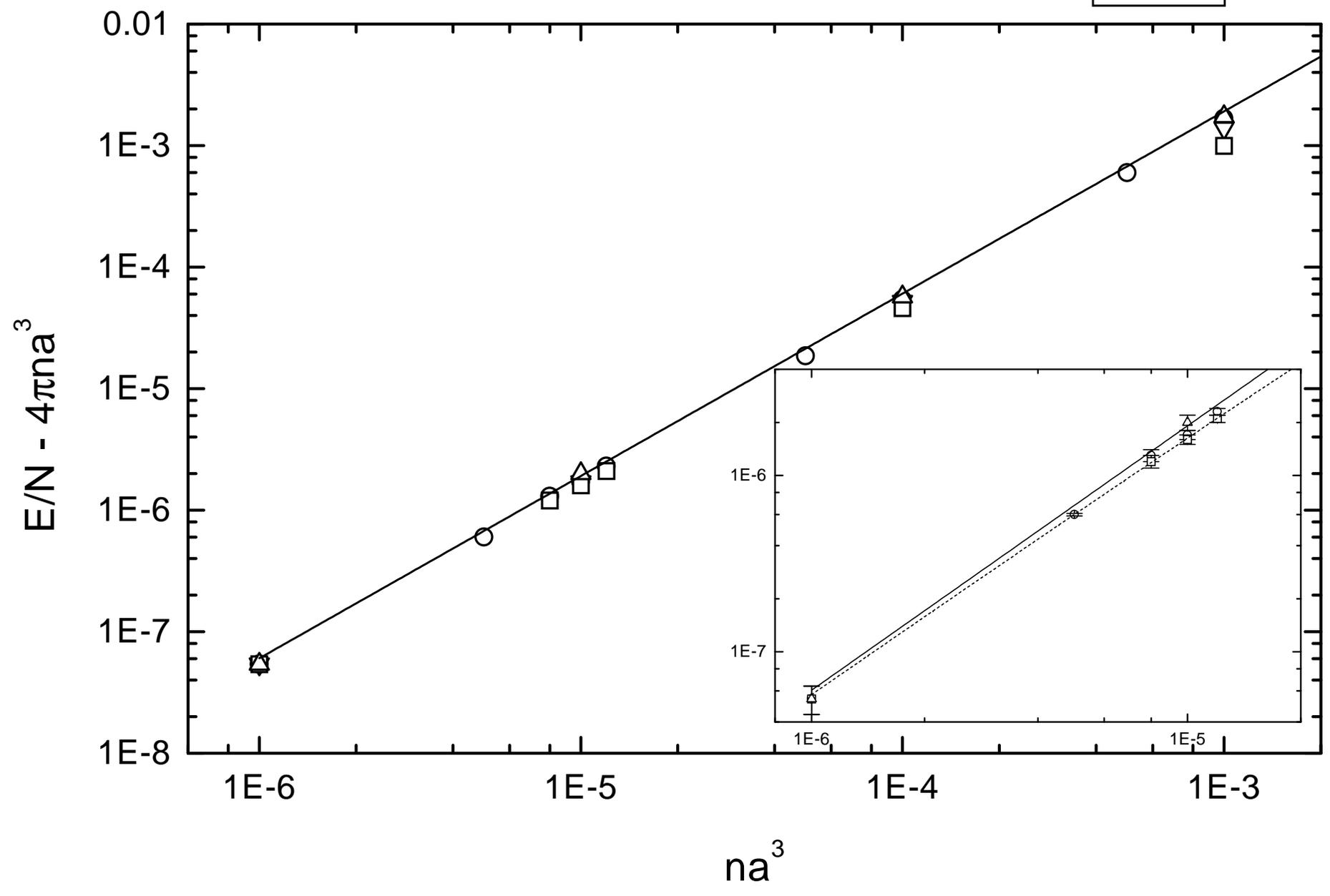
\begin{figure}
\caption{Corrections to the mean-field energy. Circles:
HS potential; down triangles: SS potential ($R=5a$); squares: SS potential ($R=10a$); up triangles:
HCSW potential; solid line: LHY correction [second term in eq. (\ref{enexp})]. Error bars are 
smaller than the size of the symbols. The inset shows the results for the HS, SS ($R=10a$), and 
HCSW potentials in the extremely low density region, the solid line is again the LHY correction, 
while the dashed line includes the logarithmic correction [second and third term in eq. 
(\ref{enexp})].}
\end{figure}

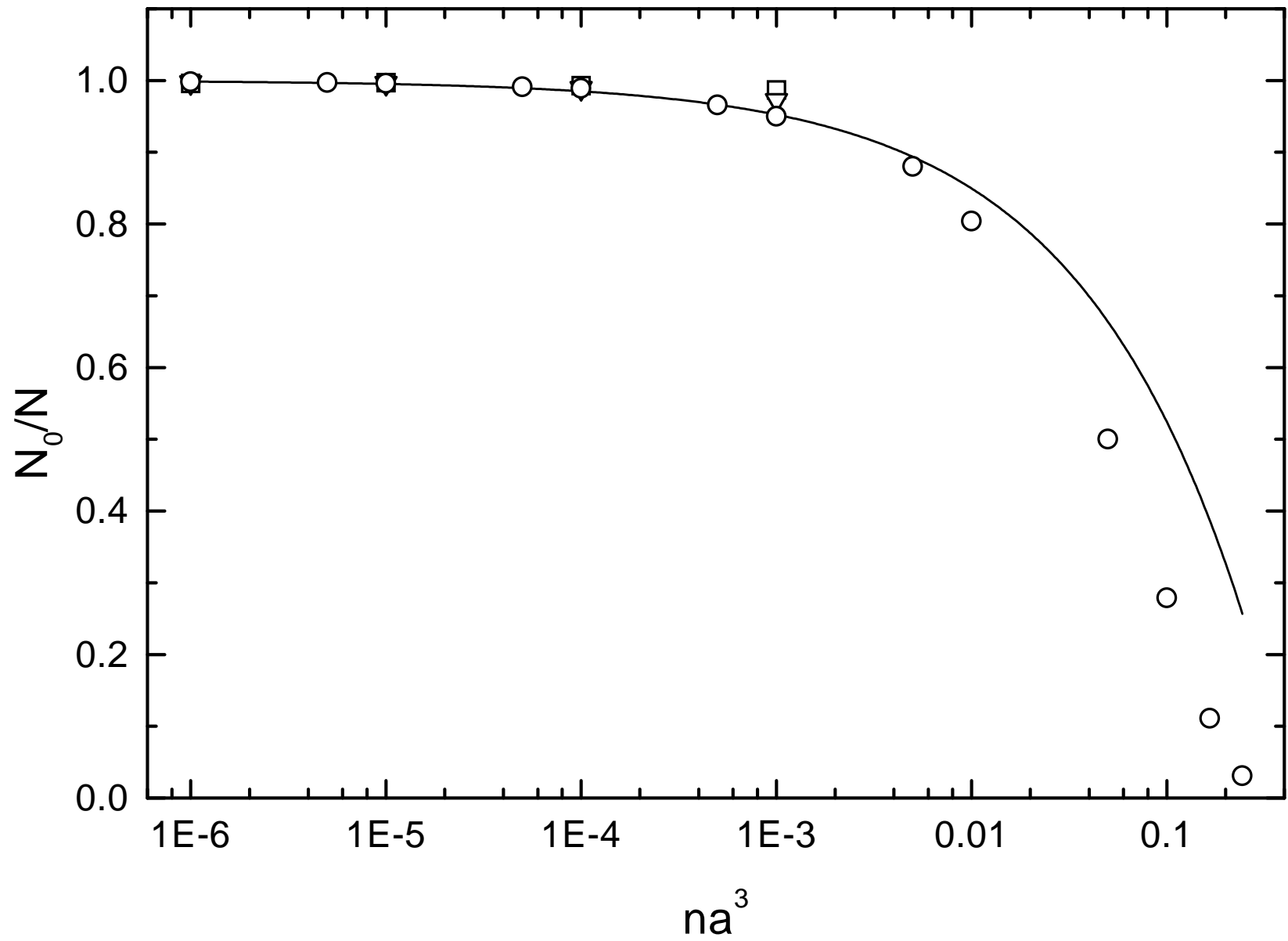
\begin{figure}
\caption{Condensate fraction as a function of the gas parameter. Circles: HS potential; down 
triangles: SS potential ($R=5a$); squares: SS potential ($R=10a$). The solid line corresponds 
to the Bogoliubov expansion (\ref{cfrac}). Error bars are smaller than the size of the symbols.}
\end{figure}

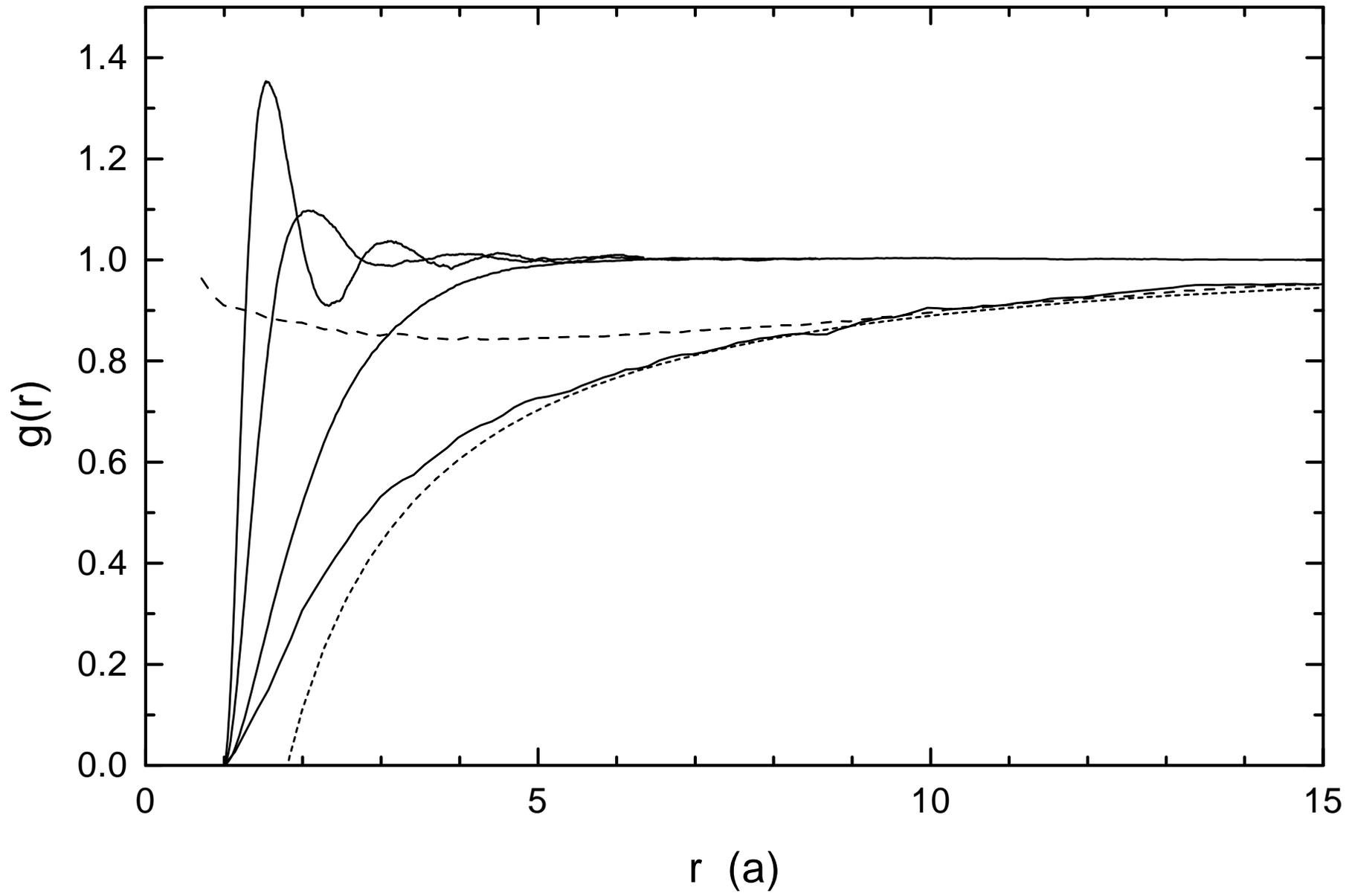
\begin{figure}
\caption{Two-body distribution function. Solid lines: hard-spheres at densities 
$na^3=$1E-4 (lowermost), 1E-2, 0.1, 0.244 (uppermost).Long-dashed line: soft-spheres 
($R=10a$) and short-dashed line: Fourier transform of the Bogoliubov static form factor 
$S(k)$ at the lowest density $na^3=$1E-4.} 
\end{figure}

\begin{table}
\caption{Energy per particle for the HS potential}
\begin{tabular}{cccc}
$na^3$ & $E/N$ ($\hbar^2/2ma^2$) & $na^3$ & $E/N$ ($\hbar^2/2ma^2$) \\ 
\tableline
1E-6 & 1.262(1)E-5 & 5E-3  & 8.154(6)E-2  \\
5E-6 & 6.343(1)E-5 & 1E-2  & 1.796(1)E-1  \\
1E-5 & 1.274(1)E-4 & 5E-2  & 1.338(1)     \\
5E-5 & 6.469(3)E-4 & 1E-1  & 3.626(7)     \\
1E-4 & 1.311(1)E-3 & 0.166 & 8.26(2)      \\
5E-4 & 6.880(4)E-3 & 0.244 & 16.7(1)      \\
1E-3 & 1.424(2)E-2 &       &              \\
\end{tabular}
\end{table}

\begin{table}
\caption{Energy per particle for the SS and HCSW potential (in units of $\hbar^2/2ma^2$)}
\begin{tabular}{cccc}
$na^3$ & $E/N$       & $E/N$        & $E/N$  \\
       & SS ($R=5a$) & SS ($R=10a$) & HCSW   \\ 
\tableline
1E-6 & 1.262(1)E-5 & 1.262(1)E-5 & 1.262(1)E-5  \\
1E-5 & 1.274(1)E-4 & 1.273(1)E-4 & 1.277(2)E-4  \\
1E-4 & 1.309(1)E-3 & 1.303(1)E-3 & 1.314(1)E-3  \\
1E-3 & 1.395(1)E-2 & 1.356(1)E-2 & 1.430(5)E-2  \\
\end{tabular}
\end{table}


\begin{references}

\bibitem{EXP} M.H. Anderson {\it et al.}, Science {\bf 269}, 198 (1995); 
K.B. Davis {\it et al.}, Phys. Rev. Lett. {\bf 75}, 3969 (1995); 
C.C. Bradley {\it et al.}, Phys. Rev. Lett. {\bf 75}, 1687 (1995).

\bibitem{REV} F. Dalfovo, S. Giorgini, L.P. Pitaevskii and S. Stringari, to appear in 
Rev. Mod. Phys..

\bibitem{BMFAN} E. Timmermans, P. Tommasini and K. Huang, Phys. Rev. A {\bf 55}, 3645 (1997);
E. Braaten and A. Nieto, Phys. Rev. B {\bf 56}, 14745 (1997), preprint cond-mat/9712041. 

\bibitem{PS} L.P. Pitaevskii and S. Stringari, Phys. Rev. Lett. {\bf 81}, 4541 (1998); 
E. Braaten and J. Pearson, Phys. Rev. Lett. {\bf 82}, 255 (1999). 

\bibitem{BMFMC} W. Krauth, Phys. Rev. Lett. {\bf 77}, 3695 (1996); P. Gr\"uter, D.M. Ceperley 
and F. Lalo\"e, Phys. Rev. Lett. {\bf 79}, 3549 (1997); S. Pearson, T. Pang and C. Chen, 
Phys. Rev. A {\bf 58}, 4796 (1998); M. Holzmann, W. Krauth and M. Naraschewski, preprint 
cond-mat/9806201. 

\bibitem{FP} A. Fabrocini and A. Polls, preprint cond-mat/9901342.

\bibitem{BOG} N.N. Bogoliubov, J. Phys. (U.S.S.R.) {\bf 11}, 23 (1947).
 
\bibitem{LHY} T.D. Lee, K. Huang and C.N. Yang, Phys. Rev. {\bf 106}, 1135 (1957). 

\bibitem{WU} T.T. Wu, Phys. Rev. {\bf 115}, 1390 (1959). 

\bibitem{BSBL} K.A. Brueckner and K. Sawada, Phys. Rev. {\bf 106}, 1117 (1957); 
{\bf 106}, 1128 (1957);
S.T. Beliaev, Zh. Eksp. Teor. Fiz. {\bf 34}, 433 (1958) [Sov. Phys. JETP {\bf 7}, 
299 (1958)]; E.H. Lieb, Phys. Rev. {\bf 130}, 2518 (1963).  

\bibitem{HP} N. Hugenholtz and D. Pines, Phys. Rev. {\bf 116}, 489 (1959).

\bibitem{LY} E.H. Lieb and J. Yngvason, Phys. Rev. Lett. {\bf 80}, 2504 (1998).

\bibitem{US} J. Boronat and J. Casulleras, Phys. Rev. B {\bf 49}, 8920 (1994).

\bibitem{HLS} J.P. Hansen, D. Levesque and D. Shiff, Phys. Rev. A {\bf 3}, 776 (1971).

\bibitem{KLV} M.H. Kalos, D. Levesque and L. Verlet, Phys. Rev. A {\bf 9}, 2178 (1974).

\bibitem{US1} J. Casulleras and J. Boronat, Phys. Rev. B {\bf 52}, 3654 (1995).

\end{references}
\end{document}